%% file: 5PM_QED.tex
\documentclass[aps,prd,nofootinbib,twocolumn,superscriptaddress,preprintnumbers]{revtex4-1}

\usepackage{amssymb}
\usepackage{amsmath}
\usepackage{epsfig}
\usepackage{hyperref}
\usepackage{breakurl}
\usepackage{bm}
\usepackage{color}
\usepackage{tikz}
\usepackage[utf8]{inputenc}
\usetikzlibrary{snakes}
\usetikzlibrary{decorations}
\usetikzlibrary{trees}
\usetikzlibrary{decorations.pathmorphing}
\usetikzlibrary{decorations.markings}
\usetikzlibrary{external}
\usetikzlibrary{intersections}
\usetikzlibrary{shapes,arrows}
\usetikzlibrary{arrows.meta}
\usetikzlibrary{calc}
\usetikzlibrary{shapes.misc}
\usetikzlibrary{decorations.text}
\usetikzlibrary{backgrounds}
\usetikzlibrary{fadings}
\usetikzlibrary{tikzmark,calc,arrows,shapes,decorations.pathreplacing}

\makeatletter
\def\simgt{\mathrel{\lower2.5pt\vbox{\lineskip=0pt\baselineskip=0pt
           \hbox{$>$}\hbox{$\sim$}}}}
\def\simlt{\mathrel{\lower2.5pt\vbox{\lineskip=0pt\baselineskip=0pt
           \hbox{$<$}\hbox{$\sim$}}}}

\makeatother

\def\spa#1.#2{\left\langle#1\,#2\right\rangle}
\def\spb#1.#2{\left[#1\,#2\right]}
\def\sand#1.#2.#3{%
\left\langle#1{\vphantom1}\right|{#2}\left|#3\right]}%
\def\sandmp#1.#2.#3{%
\left\langle#1{\vphantom1}\right|{#2}\left|#3\right]}%
\def\sandpm#1.#2.#3{%
\left[#1{\vphantom1}\right|{#2}\left|#3\right\rangle}%
\def\sandmm#1.#2.#3{%
\left\langle#1{\vphantom1}\right|{#2}\left|#3\right\rangle}%
\def\sandpp#1.#2.#3{%
\left[#1{\vphantom1}\right|{#2}\left|#3\right]}%

\def\nn{\nonumber}
\def\Section#1{\noindent {\it #1}}
\newcommand{\be}{\begin{equation}}
\newcommand{\ee}{\end{equation}}

\newcommand{\pot}{\rm pot}

\renewcommand{\imath}{\mathrm{i}}

\def\topbotatom#1{\hbox{\hbox to 0pt{$#1\bot$\hss}$#1\top$}}

\input{tikz_figs.tex}

\allowdisplaybreaks

\begin{document}
%\preprint{\preprint{CALT-TH-2021-004, FR-PHENO-2021-03, OUTP-21-03P}}

\title{Conservative binary dynamics at order $\mathcal{O}(\alpha^5)$ in electrodynamics}

\author{Zvi Bern}
\affiliation{
Mani L. Bhaumik Institute for Theoretical Physics,
University of California at Los Angeles,
Los Angeles, CA 90095, USA}

\author{Enrico Herrmann}
\affiliation{
	Mani L. Bhaumik Institute for Theoretical Physics,
	University of California at Los Angeles,
	Los Angeles, CA 90095, USA}

\author{Radu Roiban}
\affiliation{Institute for Gravitation and the Cosmos,
Pennsylvania State University,
University Park, Pz 16802, USA}

\author{Michael~S.~Ruf}
\affiliation{
Mani L. Bhaumik Institute for Theoretical Physics,
University of California at Los Angeles,
Los Angeles, CA 90095, USA}

\author{Alexander V. Smirnov}
\affiliation{Research Computing Center, Moscow State University, 119991 Moscow, Russia}
\affiliation{Moscow Center for Fundamental and Applied Mathematics, 119992 Moscow, Russia}

\author{Vladimir A. Smirnov}
\affiliation{Moscow Center for Fundamental and Applied Mathematics, 119992 Moscow, Russia}
\affiliation{Skobeltsyn Institute of Nuclear Physics of Moscow State University, 119991, Moscow, Russia}

\author{Mao Zeng}
\affiliation{Higgs Centre for Theoretical Physics, University of Edinburgh, Edinburgh, EH9 3FD, United Kingdom}

\begin{abstract}
We compute the potential-photon contributions to the classical relativistic scattering angle of two charged non-spinning bodies in electrodynamics through fifth order in the coupling. We use the scattering amplitudes framework, effective field theory, and multi-loop integration techniques based on integration by parts and differential equations. At fifth order, the result is expressed in terms of cyclotomic polylogarithms. Our calculation demonstrates the feasibility 
of the corresponding calculations in general relativity, including 
the evaluation of the encountered four-loop integrals.
\end{abstract}
   
\maketitle

%======================================================
\Section{Introduction. }
%======================================================
%
The spectacular detection of gravitational waves~\cite{LIGO} has
opened a new window on the universe.  
The expected increase in precision of up to two orders of magnitude for
the next generation gravitational wave detectors~\cite{NewDetectors} requires
commensurate advances in theoretical predictions.
Analytic perturbative approaches based on post-Newtonian
(PN)~\cite{PN} and post-Minkowskian (PM)~\cite{PM} expansions have
seen major advances in recent years---see the reviews,
e.g.\ Ref.~\cite{PNPMReview} for further details and references.  
The connection of these approaches to quantum field theory (QFT)
scattering processes has been long understood~\cite{ScatteringToGrav, DamourPM}.  
This has recently invigorated the PM approach by leveraging advances in QFT
scattering, including generalized unitarity, the double copy~\cite{BCJ} and advanced integration techniques.
The double copy~\cite{BCJ} expresses gravitational scattering amplitudes in terms
of simpler gauge-theory amplitudes, while generalized unitarity~\cite{GeneralizedUnitarity} builds
loop integrands from simpler tree-level amplitudes.  Integration by
parts (IBP) methods~\cite{IBP} allow the reduction of integrands to a
set of independent master integrals whose values can usually be
determined from differential equations~\cite{DEs,Henn}. The extraction
of classical physics from quantum scattering is greatly simplified by
basic concepts from effective field theories (EFTs), systematized for
the gravitational-wave problem in Ref.~\cite{NRGR} and applied to the
PM framework in Ref.~\cite{2PM}.  These methods have
pushed the state of the art to 4PM
$\mathcal{O}(G^4)$~\cite{3PMPRL, 3PMLong, 4PMPotential, 4PMTail}.
(See also Refs.~\cite{Dlapa:2021npj, Dlapa:2021vgp, Manohar:2022dea,
  Dlapa:2022lmu}.)  By combining perturbative, numerical
relativity~\cite{NR} and self force~\cite{self_force} results, the EOB
approach~\cite{EOB} can yield high-precision waveform templates for
both bound and unbound motion~\cite{DamourPM}.  The strikingly good
agreement between EOB-improved 4PM scattering predictions and
numerical relativity~\cite{Khalil:2022ylj} provides strong motivation
for pursuing PM calculations to ever higher orders to help match the
precision of future measurements. 

Amplitude methods efficiently solve the problem of constructing
\emph{integrands} for the gravitational two-body problem to high PM
or, equivalently, high loop orders, for example, having produced
expressions for four-point scattering at five loops in ${\cal N}=8$
supergravity~\cite{FiveLoops}.  These methods use tree-level
amplitudes as building blocks for loop-level integrands and their
efficiency derives from the physical nature of the former.  The
primary difficulty for high-order predictions is often the evaluation
of the resulting loop \emph{integrals}.
In particular, the integrals encountered in the classical gravitational two-body problem at 5PM (four-loop)
order are overwhelmingly more involved than those encountered at lower orders.

To explore possible solutions to such difficulties, we turn to the simpler
theory of electrodynamics (QED). It is a useful toy model for general
relativity (GR)~\cite{Westpfhal, Damour:1990jh, Buonanno:2000qq, KMOC, Saketh:2021sri,
Bern:2021xze} because it retains certain essential features while having far fewer integral topologies due to 
the absence of photon self-interactions. Moreover, while in the quantum theory 
the two-derivative nature of gravitational interactions leads to far more complicated 
integrals, the classical limit restricts the number of loop momenta in each diagram's numerator 
so that the gravitational integrals are of a similar complexity as the QED ones.
Our QED example demonstrates the necessary performance of the IBP
reduction for carrying out four-loop calculations. This gives us confidence
that our integration setup is applicable in GR. Here we use
\texttt{FIRE}~\cite{FIRE} and \texttt{LiteRed}~\cite{Lee:2013mka}.  We
also have written a special purpose IBP code for cross-checking
results.

In this letter, we compute potential photon contributions to the conservative QED scattering angle of two 
charged non-spinning objects to fifth order in the fine structure constant $\alpha$, leaving aside 
conservative contributions from radiation photons. In the following, we will refer to this expansion as 
the ``post-Lorentzian'' (PL) expansion, which is in direct correspondence to the PM expansion in GR.
Similarly, we refer to the analog of the PN expansion as the post-Coulombian (PC) expansion.

Unlike GR, this calculation does not exhibit divergences associated with the separation of potential and
radiation modes~\cite{TailEffect}.  The scattering angles
including radiation effects were previously found through 3PL in Refs.~\cite{Westpfhal, Saketh:2021sri, Bern:2021xze}.  

It is worth noting that QED amplitudes at small momentum transfer,
i.e.\ QED amplitudes in the classical regime, are an integral part of
the analysis of \emph{ultra-peripheral collisions} at particle
colliders and enter the relevant cross sections through interference
with nuclear S-matrix elements.  This so-called Coulomb-nuclear
interference is of current theoretical and experimental interest in
light of e.g.\ the TOTEM experiment~\cite{TOTEM:2017sdy} probing
physics down to momentum-squared transfer $|t| = 8\times 10^{-4}$~GeV$^2$ and
center-of-mass energy $\sqrt{s} = 13$~TeV.
It was first studied by Bethe in Ref.~\cite{Bethe:1958zz} and
re-analyzed from different perspectives in~\cite{Islam:1967zz}; see
Ref.~\cite{Kaspar:2011eva} for a summary of various approaches and of
the relevant cross sections.
Recent theoretical improvements include effects of excited
nuclei~\cite{Khoze:2019fhx}, an all-order-in~$\alpha$ analysis of the
leading eikonal~\cite{Kaspar:2020oih}, and an
interpretation~\cite{Petrov:2022fsu} of the data of
Ref.~\cite{TOTEM:2017sdy} in light of a new treatment of IR
divergences~\cite{Petrov:2020tnr}.
Interestingly, the momentum transfer of 2PL and 3PL terms dominates
(at low $t$) or is of the same order, respectively, as the
contribution of excited nuclei~\cite{Khoze:2019fhx}. While analyzing
them from this perspective is beyond the scope of this letter, it
would be interesting to explore the phenomenological consequences of
higher-order terms.

%======================================================
\medskip
\Section{Basic setup. }
%======================================================
%
The starting point for our calculation are quantum scattering
amplitudes, from which classical observables can be extracted via
several approaches~\cite{2PM, 3PMPRL, 3PMLong, KMOC, OtherApproaches}.
Here we use the realization that the classical elastic four-point scattering amplitude is
an appropriately-defined exponential~\cite{4PMPotential}
\begin{equation}
\label{aarelation}
	\imath \mathcal M(q) = \int_J (e^{\imath I_r(J)} - 1) \,,
\end{equation}
of the classical radial action \cite{Landau:1975pou}, $I_r(J)=\int p_r\, dr$, defined as an integral
of the radial momentum, $p_r$, over the scattering trajectory. The radial action is a function of the total angular momentum 
$J=p \, b$ of the $2\to2$ scattering process of two massive particles with center-of-mass 
momentum $p$ and impact parameter $b$;  its Fourier conjugate variable is the momentum transfer~$q$. 
The radial action (and therefore the classical limit of the amplitude) determines the scattering angle,
\begin{equation}
\label{angle}
\chi= - \frac{\partial I_r(J)}{\partial J} \,.
\end{equation}
The classical limit corresponds to large angular momentum 
$J \gg 1$ in $\hbar=1$ units, see e.g.~Refs.~\cite{2PM,3PMLong},
and translates to the hierarchy of scales 
$s,m_1^2,m_2^2 \sim J^2 |t| \gg |t| = |q|^2.$
The phase space splits into two regions 
\begin{equation}
\label{eq:hard_soft}
\text{hard:}\quad\ell\gg |q| \,, \qquad\text{soft:}\quad      \ell\sim |q| \,.
\end{equation}
Classical physics is captured by the soft region. At $L$ loops, it is contained in the terms proportional to $|q|^{L-2} \ln |q|$ and $|q|^{L-2}$ for even and odd $L$, respectively. 
To refine these contributions, we identify the potential and radiation subregions~\cite{Beneke:1997zp}, characterized by a small velocity $v$: 
\begin{equation} 
\label{eq:ClassicalRegions}
\null \hskip -.35 cm
\text{potential: } \ell\sim(v,1)|q|\,,\qquad
\text{radiation: } \ell\sim(v,v)|q| \,. 
\end{equation} 
Here we focus on the contribution where all photon loop
momenta $\ell_i$ are in the potential region. As in GR, the potential
region does not account for all conservative effects, which, starting at 4PM/4PL~\cite{4PMTail}, 
also require the inclusion of radiation modes. 
At 5PL, we leave such contributions to future work. However, in contrast to the
gravitational case, the potential-region contribution in QED  gives
rise to a well-defined local classical potential due to the absence of
nonlinearities of the field equations and the associated absence of
tail effects~\cite{TailEffect}. Nonetheless, we use dimensional regularization,
setting $D=4-2\epsilon$, to handle divergences in intermediate expressions.

\begin{figure}[tb]
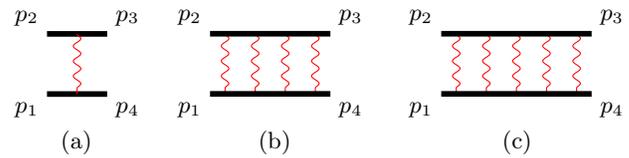

\treetPhoton \hskip .3 cm \PlanarTripleLadder  \hskip .3 cm \PlanarQuadLadder \\[-2pt]
\hskip -.5 cm  (a) \hskip 2.1 cm (b) \hskip 2.7 cm (c) \hskip 1. cm 
 \caption{Sample Feynman diagrams describing the scattering of two (macroscopic) charges at (a) leading, (b) fourth , and (c) fifth order in $\alpha$ in an all-outgoing momentum convention.}
\label{fig:egFeynmanDiags}
\end{figure}

%======================================================
\medskip
\Section{QED and loop integrands. }
%======================================================
%
To describe two electrically-charged classical spinless compact objects of mass $m_1$ and $m_2$ 
at scales much larger than their size, we use
the minimally-coupled scalar QED Lagrangian in $R_\xi$ gauge,
\begin{align}
\label{eq:Lagrangian}
\begin{split}
\hspace{-.4cm}
\mathcal{L}  = & {-}\frac{1}{4} F^2_{\mu \nu}  
			{-} \frac{1}{2\xi} \left(\partial_\mu A^\mu\right)^2 
			{+} \sum^2_{i=1}  \Big[ \left| D_\mu \phi_i\right|^2 {-} m^2_i \left| \phi_i\right|^2 \Big] \,,
\hspace{-.4cm}			
\end{split}		      
\end{align}
where the covariant derivative $D_\mu = \partial_\mu - \imath \, e\, Q_i\,  A_\mu$ contains the photon field $A_\mu(x)$ and the electric charge $Q_i$ for each object measured in terms of the elementary charge $e$, related to the fine structure constant $\alpha = e^2/(4\pi) \simeq 1/137$. Below it will be convenient to define an effective small coupling  $\alpha_{\mathrm{eff}}={\alpha \,Q_1Q_2}/{J}$ in terms of the charges of the scattering objects.   The Maxwell field strength is $F_{\mu \nu} = \partial_\mu A_\nu - \partial_\nu A_\mu$.  The Feynman vertices of scalar electrodynamics used in the calculation are derived from the Lagrangian in Eq.~(\ref{eq:Lagrangian}) using standard procedures.\footnote{Alternatively, in order to reproduce the conservative potential contribution, one could integrate out the photon field at tree-level to produce a Fokker-type action~\cite{Kerner,Damour:1990jh}.}   This allows us to construct  QED amplitudes  via Feynman diagrams to higher orders in perturbation theory.  At leading order in $\alpha$ only a single diagram contributes. At order $\alpha^5$ there are a total of 1536 relevant Feynman diagrams, which we organize into 23 graphs containing only cubic vertices. Sample diagrams at various orders in perturbation theory are shown in Fig.~\ref{fig:egFeynmanDiags}.  The primary calculation is performed in Feynman gauge $\xi = 1$, but 
we verified the $\xi$-independence of the amplitudes at one numerical kinematic point.

%%%%%%%%%%%%%%%%%%% FIGURE %%%%%%%%%%%%%%%
%%%%%%%%%%%%%%%%%%%%%%%%%%%%%%%%%%%%%%%%%%

%======================================================
\medskip
\Section{Soft expansion, integral reduction and integration. }
%======================================================
%
With the $L$-loop integrand at hand we extract terms that can contribute to the classical limit by expanding in the soft region (\ref{eq:hard_soft}) to $L^{{\rm th}}$ order. 
Starting from the $p_i$ in Fig.~\ref{fig:egFeynmanDiags}, special variables~\cite{PRZ},
\begin{align}
\hspace{-.5cm}
\overline{p}_1 = {-}p_1 {+} \frac{q}{2} = p_4{-}\frac{q}{2}\,,  \quad
\overline{p}_2 = {-}p_2 {-} \frac{q}{2}  = p_3{+}\frac{q}{2}\,,
%p_3 {=} \overline{p}_2 {-}  \frac{q}{2} \,, \ \
%p_4 {=} \overline{p}_1 {+}  \frac{q}{2} \,,
\hspace{-.5cm}
\end{align}
which satisfy $\overline{p}_1 \cdot q = \overline{p}_2 \cdot q = 0,\, \overline{p}_i^2= \overline{m}_i^2 $ simplify the analysis. After the soft expansion, the dependence on $q^2$ and $\overline{m}_i$ is trivially fixed by scaling arguments. Therefore all integrals are functions of the single kinematic variable,
\begin{align}
\label{defYvsXvsSigma}
y 	:= \frac{\overline{p}_1 \cdot \overline{p}_2}{\overline{m}_1  \overline{m}_2} 
	= \frac{1+x^2}{2x} 
	= \sigma + \mathcal O (q^2)\,, 
\end{align}
where the $x$ parametrization is useful to simply 
the resulting expressions, and $\sigma{=}p_1{\cdot} p_2/(m_1m_2) {=} (1{-}v^2)^{-1/2}$.

The soft expansion yields many terms, some with higher-rank tensor integrals and 
higher-power matter propagators. 
Yet, the key factor characterizing the difficulty of the calculation is not the number of
integrals but rather their ``IBP complexity'', defined for each of them as
the number of irreducible numerators plus the number of ``dots''
corresponding to doubled and higher powers of propagators. For QED at
$L$-loops, one obtains integrals with up to rank $L$ numerators and up
to $L$ additional dots for a maximum IBP
complexity of $2L$.  Remarkably, similar counting in the classical
limit shows that the maximum IBP complexity in GR is identical to
that of QED. We should mention, however, that besides the appearance of many more
diagram families in GR, they require more effort to evaluate
at the same complexity ranking.  In contrast, the ladder diagrams of
GR can be evaluated using precisely the same setup as for QED.

The integral of the soft-expanded integrand is subsequently reduced to a combination
of master integrals via IBP.  These integrals are naturally
organized in terms of families of diagrams with only cubic vertices and their
contractions (contact diagrams).  Different families share many common
contact integrals so that we find it convenient to organize the
computation in terms of a single set of ``global'' master integrals
shared across all families. The global master integrals are evaluated
via differential equations~\cite{DEs}, following the steps described
in Ref.~\cite{PRZ}. In particular, we use an implementation of Lee's
algorithm~\cite{Lee:2014ioa}, together with various software packages~\cite{Prausa:2017ltv} to find a basis of
master integrals $\vec{\mathcal{I}}(x,\epsilon)$ that brings the
corresponding differential equation to canonical form~\cite{Henn},
\begin{equation}
\label{eq:can_diff_eq}
	\partial_x \vec{\mathcal{I}}(x,\epsilon) =
	\epsilon\sum_{w\in\mathbb{W}}  w(x)\,  A_w \, \vec{\mathcal{I}}(x,\epsilon)\,,
\end{equation}
with rational matrices $A_w$ and logarithmic kernels $\mathbb{W}$.

To fully specify the solutions we supply boundary conditions in the near-static 
limit $x\to 1$ (equivalent to $v\to0$) by expanding the master integrals in velocity 
as explained in Ref.~\cite{PRZ}. Another constraint is analyticity at $v\to0$. 
The form of the resulting solutions are power series in $\epsilon$ whose 
coefficients are generically generalized polylogarithms~\cite{Goncharov:1998kja} in $x$.

%======================================================
\medskip
\Section{Potential-region scattering angles through $\mathcal{O}(\alpha^4)$. }
%======================================================
%
Eqs.~\eqref{aarelation} and \eqref{angle} give the potential-region contributions to the scattering angles through fourth order as
\begin{align}
\label{eq:cons_angles_J1234}
\chi^{\rm 1PL}_{\pot} &=
%%%%% begin : chiPaper[1]
%\frac{\alpha\, Q_1 Q_2}{J}
\alpha_{{\rm eff}}\frac{-2 \sigma}{\sqrt{\sigma^2-1}}
%%%%% end : chiPaper[1]
 \,, \nonumber \\
\chi^{\rm 2PL}_{\pot} &= 
%%%%% begin : chiPaper[2]
\alpha_{{\rm eff}}^2
\frac{\pi}{2 \sqrt{1+2\nu(\sigma-1)}}
%%%%% end : chiPaper[2]
  \,, \nonumber \\
  \chi^{\rm 3PL}_{\pot} &=
%%%%% begin : chiPaper[3]
\alpha_{{\rm eff}}^3
\frac{{-}2\sigma (2\sigma^2{-}3)+4\nu(\sigma{-}1)(\sigma^3{+}3\sigma^2{-}3)}
       {3\, (1+2\nu (\sigma-1))\, (\sigma^2-1)^{3/2}} 
%%%%% end : chiPaper[3]
 \,, \nonumber \\
 \chi^{\rm 4PL}_{\pot} &= 
%%%%% begin : chiPaper[4]
\alpha_{{\rm eff}}^4
\frac{3 \pi}{8 (1+2\nu(\sigma-1))^{3/2}} \biggl[ 1 
 \\
& 
\hskip -.8cm + \frac{\nu}{2(\sigma^2{-1})} 
	\bigg\{ 
	 3 \sigma^4{-}11 \sigma ^3{+}3 \sigma ^2{+}\sigma {+}14{-}\frac{7 \sigma ^2{-}1}{\sigma ^3}
	\nonumber \\ & 	
\hskip -.8cm	
\hskip .15cm 
+
	2\left(3 \sigma^3{-}4 \sigma^2{+}9 \sigma{-}4\right) \! \frac{\log (x)}{\sqrt{\sigma ^2{-}1}}
	{+}
	(3 \sigma^2{+}1) \!\left[\frac{\log (x)}{\sqrt{\sigma^2{-}1}}\right]^2
	\!\!\bigg\} \bigg]
%%%%% end : chiPaper[4]
\nonumber
\,,
\end{align}
where ${\nu=m_{1}m_2}/{(m_1+m_2)^2}$ is the symmetric mass ratio. In Refs.~\cite{Saketh:2021sri, Bern:2021xze}, the scattering angles, including radiative effects, were presented through $\alpha^3$.  The $\alpha^4$ potential-region contribution is new. For completeness, we have also computed the conservative radiation contribution to the scattering angle using the methods described in Ref.~\cite{4PMTail} and provide the result in the ancillary file \cite{AttachedFile}. Interestingly,  $\chi^{\rm 4PL}_{\pot}$ is related by a derivative with respect to~$\log (x)$ (holding other appearances of $\sigma$ fixed) to the energy-loss at 3PL~\cite{Saketh:2021sri, Bern:2021xze} in the $Q^3_1Q^3_2$ charge sector
\begin{align}
\hspace{-.4cm}
\Delta E^{Q^3_1Q^3_2}_{ {\rm c.m.}} & \sim
 \left(3 \sigma^3 {-} 4 \sigma ^2 {+} 9 \sigma {-} 4\right) +\left(3 \sigma^2 {+} 1\right)\! \frac{\log(x)}{\sqrt{\sigma^2 - 1}}\,.
 % \!\Bigg]\,. \nn
 \hspace{-.4cm}
\end{align}
In GR, the 3PM energy loss is related to the divergent part of the 4PM tail contribution~\cite{3PMEnergyLoss} which 
in the full expression cancels against
the divergent part of the potential-region scattering angle. In contrast, the relation appears to hold for the 
finite part in QED but not in GR. It would be interesting to investigate this further. 

\medskip
\Section{Scattering angles at $\mathcal{O}(\alpha^5)$. }
Compared to lower orders, where analogous GR calculations are available, our four-loop result is at a previously unexplored order in the PM/PL expansion and warrants some further details. 

The complete 5PL amplitude is composed of 1536 Feynman diagrams of which we have to evaluate 213, while the rest can be obtained by crossing. For the purpose of the integrand reduction via integration by parts, we organize the integrand into 23 ladder-type diagrams with cubic vertices by multiplying and dividing by any missing propagators. The soft expansion is straightforward. After applying IBP reduction to the soft expanded four-loop integrand we obtain a set of master integrals for each family.  Removing redundant master integrals between the 23 families leaves 1107 global master integrals. Here, we again bring the master integral differential equations into canonical form, see Eq.~(\ref{eq:can_diff_eq}).

The logarithmic kernels for the differential equation at four loops are explicitly given by:
\begin{align}
\label{eq:eg_cyclotomic_kernels_DE}
\begin{split}
\hspace{-.3cm}
\mathbb{W} & = \left\{
				\frac{1}{x}, 
				\frac{1}{1{+}x}, 
				\frac{1}{x{-}1},
				\frac{2x}{1{+}x^2},
				\frac{1+2x}{1{+}x{+}x^2}\,,
				\frac{2x-1}{1{-}x{+}x^2}
			\right\}
			\\
		& {:=}\left\{
				f^0_0, \ 
				f^0_2,  \quad
				f^0_1, \quad 
				2f^1_4, \ \ 
				2f^1_3+f^0_3,\ 
				2f^1_6-f^0_6
			\right\}\,,
			\hspace{-.3cm}
\end{split}			
\end{align}
and re-expressed via cyclotomic kernels $f^i_j{:=} x^i/\Phi(j,x)$; $\Phi(j,x)$ being the $j^{{\rm th} }$ cyclotomic polynomial.
Therefore, the solutions of the differential equations are naturally written in terms of cyclotomic harmonic polylogarithms (CPL's), originally introduced in Ref.~\cite{Ablinger:2011te}, and have appeared in different contexts, see e.g.~Ref.~\cite{Ablinger:2018zwz},
\begin{align}
	C_{a_1,\dots, a_n}^{b_1,\dots, b_n}(x) &= \int_0^x \mathrm{d}z \, f_{a_1}^{b_1}(z)\,  C_{a_2,\dots  ,a_n}^{b_2, \dots , b_n}(z)\,, 
\end{align}
with $C^{0}_{0}(x) :=\log(x)$.

The boundary conditions are evaluated in the near-static limit $x\to 1$. Analyticity in this limit provides 814 boundary conditions.
The remaining 293 conditions correspond to appropriately normalized scalar integrals which are fixed by explicitly expanding in the near-static limit, as explained in Ref.~\cite{PRZ}.

%======================================================
%
%\medskip
%\paragraph{Amplitude}  
%
%======================================================

Inserting the master-integral values into the IBP-reduced integrand yields the potential part of the classical amplitude for the scattering of two charged particles,
\begin{align}
\begin{split}
\hspace{-.2cm}
\mathcal{M}_5 =&
%%%%% begin : M5Paper
- J^5\alpha_{\mathrm{eff}}^5\, |q|^2\log(|q|^2)
		\bigg[{-}4 \sigma{} \left(15{-}20\sigma^2{+}8\sigma^4\right) \\
		&\quad + \sum_{k=1}^{12}\left(\nu\, r_k^{(1)} + \nu^2\, r_k^{(2)}\right) f_k \bigg]	
%%%%% end : M5Paper
+\text{iteration,}
\hspace{-1cm}
\end{split}
\label{M5}
\end{align}
where the result is organized in terms of the symmetric mass ration
$\nu$.  The ``iterations'' are dictated by the amplitude-action
relation in Eq.~(\ref{aarelation})~\cite{4PMPotential} and we verified
their connection with lower-loop amplitudes.  We also introduced  a basis 
of transcendental functions $f_k$ and algebraic coefficients $r_k^{(i)}$, 
\begin{align}
	r^{(1)}_{1} ={}& 
%%%%% begin : rPaper[1,1]
	\frac{15}{\sigma ^2}
	-208 \sigma ^6+128 \sigma ^5-625 \sigma ^4-320 \sigma ^3+705 \sigma ^2\nonumber\\
	&+240 \sigma +65
	%-\frac{(\sigma -1) \left(208 \sigma ^7+80 \sigma ^6+705 \sigma ^5+1025 \sigma ^4+320 \sigma ^3+80 \sigma ^2+15 \sigma +15\right)}{\sigma ^2}
%%%%% end : rPaper[1,1]
	\,, \nn\\
	r^{(1)}_{2} ={}&
	%
%%%%% begin : rPaper[1,2]
	\sqrt{\sigma ^2-1} \left[
	-\frac{60 \left(5 \sigma ^2-1\right)}{\sigma ^3} -80 \sigma  {} \left(16 \sigma ^2+23\right)
	\right]
	%-\frac{20 \sqrt{\sigma ^2-1} \left(64 \sigma ^6+92 \sigma ^4+15 \sigma ^2-3\right)}{\sigma ^3}
%%%%% end : rPaper[1,2]
	\,,\nn \\
	r^{(1)}_{3} ={}&
	%
%%%%% begin : rPaper[1,3]
	\frac{90 \left(6 \sigma ^2-1\right)}{\sigma ^4}-10 \left(350 \sigma ^2+319\right)
%%%%% end : rPaper[1,3]
	\,,\nn \\
	%-\frac{10 \left(350 \sigma ^6+319 \sigma ^4-54 \sigma ^2+9\right)}{\sigma ^4}
	%
	r^{(1)}_{4} ={}& 
%%%%% begin : rPaper[1,4]
	-\frac{5760 \sigma }{\sqrt{\sigma ^2-1}}
%%%%% end : rPaper[1,4]
	\,,\nn \\
	r^{(1)}_{5} ={}& 
%%%%% begin : rPaper[1,5]
	120\left(\sigma ^2-1\right)^{3/2} \left(2 \sigma ^2-1\right)
	%30 \left(\sigma ^2-1\right)^{3/2} \left(2 \sigma ^2-1\right)
%%%%% end : rPaper[1,5]
	\,,\nn \\
	r^{(1)}_{8} ={}&
%%%%% begin : rPaper[1,8]
	120 \left(\sigma ^2-1\right) (\sigma^2+\sigma-1)
%%%%% end : rPaper[1,8]
	%240 \left(\sigma ^2-1\right)^2
	\,,\nn\\
	r^{(1)}_{9} ={}&r^{(1)}_{12}=  
%%%%% begin : rPaper[1,9,12]
	240 (\sigma^2-1)^2
%%%%% end : rPaper[1,9,12]
	\,,\nn\\
	%-30 \left(\sigma ^2-1\right) \left(\sigma ^2+2 \sigma -1\right)
	%
	r^{(1)}_{11} ={}& 
%%%%% begin : rPaper[1,11]
	120(\sigma^2-1)(\sigma^2+2\sigma-1)
%%%%% end : rPaper[1,11]
	%240 \left(\sigma ^2-1\right)^2
	\,,\nn\\
		r^{(1)}_{6} ={}&r^{(1)}_{7} = r^{(1)}_{10} ={}
%%%%% begin : rPaper[1,6,7,10]
        0
%%%%% end : rPaper[1,6,7,10]
         \,, \nn \\
%\end{align}
%%%%%%%%%%%%%%%%%%%%%%%%%%%%%%%%%%%%%%%%%%%%%
%\begin{align}
	r^{(2)}_{1} ={}& 
%%%%% begin : rPaper[2,1]
	\frac{405 \sigma  \left(15{-}44 \sigma ^2\right)}{16 \left(1{-}4 \sigma ^2\right)^2}
	- \frac{15 \left(10 \sigma ^2{+}2 \sigma {-}3\right)}{\sigma ^3}\nn\\% +\frac{45-30 \sigma  (5 \sigma {+}1)}{\sigma ^3}\\
	&+\frac{{-}2048 \sigma ^7{+}6656 \sigma ^6{+}17872 \sigma ^5{+}20000 \sigma ^4} {16} \nn\\
	&+\frac{{-}7740 \sigma ^3{-}22560 \sigma ^2{-}6635 \sigma {-}2080}{16} 
%%%%% end : rPaper[2,1]
	\,,\nn\\
	r^{(2)}_{2} ={}& 
%%%%% begin : rPaper[2,2]
	\sqrt{\sigma ^2{-}1} \biggl[
	\frac{45 \left(1232 \sigma ^4{-}1168 \sigma ^2{+}287\right)}{16 \left(4 \sigma ^2-1\right)^3}\nn\\
	&
	+\frac{30 \left(20 \sigma ^3{-}9 \sigma ^2{-}4 \sigma {+}3\right)}{\sigma ^4} \nn\\
	&
	+\frac{5}{16} \left(1776 \sigma ^4{+}8192 \sigma ^3{+}10820 \sigma ^2{+}11776 \sigma {+}3223\right)
	\biggr]
%%%%% end : rPaper[2,2]
	\,,\nn\\
	r^{(2)}_{3} ={}& 
%%%%% begin : rPaper[2,3]
	-\frac{30 \left(16 \sigma ^4+36 \sigma ^3-11 \sigma ^2-6 \sigma +3\right)}{\sigma ^5} \nn\\
	&+20 \left(212 \sigma ^3+350 \sigma ^2+328 \sigma +319\right)
%%%%% end : rPaper[2,3]
	\,,\nn\\
	r^{(2)}_{4} ={}&
%%%%% begin : rPaper[2,4]
	\frac{2880 (\sigma +1) (3 \sigma +1)}{\sqrt{\sigma ^2-1}}
%%%%% end : rPaper[2,4]
	\,,\nn\\
	r^{(2)}_{6} ={}&
%%%%% begin : rPaper[2,6]
	480 \left(\sigma ^2-1\right)^{3/2} \left(2 \sigma ^2-1\right)
%%%%% end : rPaper[2,6]
	\,,\quad 
	\nn\\
	r^{(2)}_{7} ={}&
%%%%% begin : rPaper[2,7]
	45 \sigma  \left(\sigma ^2-1\right)^{5/2}
%%%%% end : rPaper[2,7]
	,\nn\\
	r^{(2)}_{9} ={}& 
%%%%% begin : rPaper[2,9]
	- \! 480 \left(\sigma ^2-1\right) \left(\sigma ^2-\sigma -1\right)
%%%%% end : rPaper[2,9]
	\,,\nn\\
	r^{(2)}_{10} ={}&
%%%%% begin : rPaper[2,10]
	- \! 135 \left(\sigma ^2-1\right)^2
%%%%% end : rPaper[2,10]
	\,,\nn\\
	r^{(2)}_{12} ={}& 
%%%%% begin : rPaper[2,12]
	- \! 480 \left(\sigma ^2-1\right) \left(\sigma ^2-2 \sigma -1\right)
%%%%% end : rPaper[2,12]
	,\nn\\
	r^{(2)}_{5} ={}&r^{(2)}_{8} =r^{(2)}_{11} =
%%%%% begin : rPaper[2,5,8,11]
	0
%%%%% end : rPaper[2,5,8,11]
	\,.
\end{align}
The transcendental functions have compact representations in terms of CPL's  in the variable $x$ 
\begin{align}
f_{1} ={}& 
%%%%% begin : fPaper[1]
    1
%%%%% end : fPaper[1]
 \,, ~
 %\nn\\
%
f_{2} ={}%& 
%%%%% begin : fPaper[2]
   C_{0}^{0}(x)
%%%%% end : fPaper[2]
\,,~%\nn \\
f_{3} ={}%& 
%%%%% begin : fPaper[3]
   C_{0,0}^{0,0}(x)
%%%%% end : fPaper[3]
%
 \,,~
 %\nn\\
f_{4} ={}%& 
%
%%%%% begin : fPaper[4]
  C_{0,0,0}^{0,0,0}(x)
%%%%% end : fPaper[4]
%
\,, \nn\\
f_{5} ={}& 
%%%%% begin : fPaper[5]
  -C_{1,0}^{0,0}(x)+C_{2,0}^{0,0}(x)+\frac{\pi ^2}{4}
%%%%% end : fPaper[5]
\,,\nn \\
f_{6} ={}& 
%%%%% begin : fPaper[6]
 -C_{2,0}^{0,0}(x)+C_{4,0}^{1,0}(x)-\frac{\pi ^2}{16}
%%%%% end : fPaper[6]
\,,\nn \\
f_{7} ={}& 
%%%%% begin : fPaper[7]
  C_{3,0}^{0,0}(x){+}2 C_{3,0}^{1,0}(x){+}C_{6,0}^{0,0}(x){-}2 C_{6,0}^{1,0}(x){+}\frac{\pi ^2}{6}
%%%%% end : fPaper[7]
\,,\nn \\
f_{8} ={}& 
%%%%% begin : fPaper[8]
-C_{0,1,0}^{0,0,0}(x)+C_{0,2,0}^{0,0,0}(x)+\frac{\pi ^2}{4}  C_{0}^{0}(x)+\frac{7 \zeta _3}{2}
%%%%% end : fPaper[8]
\,,\nn \\
f_{9} ={}& 
%%%%% begin : fPaper[9]
-C_{0,2,0}^{0,0,0}(x)+C_{0,4,0}^{0,1,0}(x)-\frac{ \pi ^2}{16} C_{0}^{0}(x)-\frac{21 \zeta _3}{16}
%%%%% end : fPaper[9]
\,,\nn \\
f_{10} ={}&
%%%%% begin : fPaper[10]
 C_{0,3,0}^{0,0,0}(x)+2 C_{0,3,0}^{0,1,0}(x)+C_{0,6,0}^{0,0,0}(x)-2 C_{0,6,0}^{0,1,0}(x)
  \nonumber\\
&+\frac{1}{6} \pi ^2 C_{0}^{0}(x)+\frac{28 \zeta _3}{9}
%%%%% end : fPaper[10]
\,,\nn \\
f_{11} ={}& 
%%%%% begin : fPaper[11]
-C_{1,0,0}^{0,0,0}(x)+C_{2,0,0}^{0,0,0}(x)-\frac{7 \zeta _3}{4}
%%%%% end : fPaper[11]
\,, \nn \\
f_{12} ={}& 
%%%%% begin : fPaper[12]
-C_{2,0,0}^{0,0,0}(x)+C_{4,0,0}^{1,0,0}(x)+\frac{21 \zeta _3}{32}
%%%%% end : fPaper[12]
\,.\label{eq:fiCPL}
\end{align}
Remarkably, the amplitude \eqref{M5}
does not depend on the full cyclotomic alphabet, but only on the
subset
\begin{align}
	\mathbb{W}^\prime=\left\{\frac{1}{x},\frac{x}{1{-}x^2},\frac{x-1}{(x{+}1) \left(1{+}x^2\right)},\frac{1-x^2}{1{+}x^2{+}x^4}\right\}\,.
\end{align}
Note that the last letter appears only in the $\nu^2$ sector of the amplitude, while the second letter does not appear there.
The transcendental constants in Eq.~\eqref{eq:fiCPL} are chosen such that the expansion of the functions around the static point $x=1$ only involves rational numbers. Therefore the post-Coulombian expansion of the amplitude
is manifestly free of these constants and it is natural to conjecture that this property holds to all orders.
This is in contrast to GR where $\pi^2$ is present at 4PM and is closely tied to the appearance of elliptic integrals.  For the particular combination of CPL's in the classical amplitude we find that there exists an expression in terms of classical polylogarithms with \emph{real} arguments. The expression can be found in the ancillary files to this letter~\cite{AttachedFile}.

The ${\cal O}(\alpha^5)$ angle follows from Eqs.~\eqref{M5}, \eqref{aarelation} and \eqref{angle}:
\begin{align}
\label{5PMangle}
	\chi^{\rm 5PL}_{\pot}  ={}&
%%%%% begin : chiPaper[5]
 \alpha_{\rm eff}^5 \frac{1}{30\,(1+2\nu(\sigma-1))^{2}\,(\sigma^2-1)^{5/2}} \\
& \hskip -1cm\times 
\left[{-}4\sigma\left(15{-}20\sigma^2{+}8\sigma^4\right)
         { +} \sum_{k=1}^{12}\left(\nu\, r_k^{(1)}+\nu^2\,  r_k^{(2)}\right) f_k \right]
%%%%% end : chiPaper[5]
.\nonumber
\end{align}
For both the 4PL and 5PL potential-region scattering angles we performed a number of nontrivial checks.
We verified the independence of the result on the gauge parameter $\xi$ after IBP reduction 
to a master-integral basis, checking much of the integrand and integral tables.

We have also computed the scattering angle to fourth post-Coulombian order using the Fokker-type 
Lagrangian of Wheeler-Feynman electrodynamics~\cite{Kerner} from Eq.~(4.1) of Ref.~\cite{Damour:1990jh}, and 
applied a variant of their proposed order-reduction procedure to eliminate higher time derivatives.\footnote{We thank Justin Vines for discussions on this topic and for sharing an unpublished result at 3PC order~\cite{VinesUnpublished}.}
For the overlapping terms, we find complete agreement with the velocity expansion of our PL expressions in the near-static limit. In particular we match the expansion of the 5PL angle,
\begin{align}
	\chi^{ \rm 5PL}_{\rm pot}={}
	& \alpha_{{\rm eff}}^5\left[
		-\frac{2}{5 v^5}
		+\frac{4}{3 v^3}
		+\frac{2 (8 \nu -3)}{3 v}
		+\frac{8}{9} \nu  (5-18 \nu ) v 
		\right.\nonumber\\
		&\left. 
		+\nu\left(\frac{80 \nu ^2}{3}-\frac{532 \nu }{27}+\frac{226  }{45}\right)v^3+\mathcal{O}(v^5)
	\right].
\end{align}
Notice that the $\mathcal{O}(\alpha^5)$ potential starts contributing only at $\mathcal{O}(v^3)$, 
while $\mathcal{O}(v^{n\le 1})$ are fixed by lower PL orders.

The probe limit,  $\nu\to 0$, in which one mass is much smaller than the other, provides another important check.
In this limit the angle has a simple expression~\cite{QEDProbe,Landau:1975pou} to all orders in $\alpha$: 
\begin{equation}
  \hskip -.2 cm
     \chi^{(0)}= -\pi + \frac{2}{\sqrt{1-\alpha_{\mathrm{eff}}^2 } } \arctan\biggl[
		\frac{ \sqrt{\sigma^2{-}1}}{\sigma} \frac{\sqrt{1{-}\alpha_{\mathrm{eff}}^2}}{\alpha_{\mathrm{eff}}}\biggr] \, .
\end{equation}
Expanding in small $\alpha_{\mathrm{eff}}\ll1$, we obtain 
\begin{align}
\chi^{(0)}  = & - \alpha_{\mathrm{eff}} \frac{2 \sigma }{\sqrt{\sigma ^2-1}}
		  +\alpha_{\mathrm{eff}}^2 \frac{\pi}{2}
		  +\alpha_{\mathrm{eff}}^3 \frac{2\sigma \left(2 \sigma ^2-3 \right)}{3 \left(\sigma ^2-1\right)^{3/2}} 
		  \nn \\ &  \hspace{-.8cm}
		  + \alpha_{\mathrm{eff}}^4 \frac{3 \pi}{8}
		  - \alpha_{\mathrm{eff}}^5  \frac{2 \sigma  \left(8 \sigma ^4-20 \sigma ^2+15\right)}{15 \left(\sigma ^2-1\right)^{5/2}}
		  + \mathcal{O}(\alpha_{\mathrm{eff}}^6) \,,
\end{align}
in agreement with the probe-limit of the explicit PL
results in Eqs.~\eqref{eq:cons_angles_J1234} and \eqref{5PMangle}.

We note that $\chi^{ \rm 5PL}_{\rm pot}$ exhibits singularities at
$\sigma=0,\pm 1/2$ which lie outside of the physical scattering region
$1<\sigma$. Similar poles are present in the complete 4PM results in
GR~\cite{4PMPotential, 4PMTail, Dlapa:2021npj, Dlapa:2021vgp,
  Manohar:2022dea, Dlapa:2022lmu}. It would be interesting to
investigate their fate after the analytic continuation to the bound
regime~\cite{B2B}. QED offers a cleaner environment to study this issue
due to the absence of the tail effect.

%======================================================
%
\medskip
\Section{Conclusions. }
%
%======================================================
%
In this letter we studied the scattering of two classical charges through fifth order in the fine 
structure constant $\alpha$. While our primary objective was to connect to the rapid progress 
in the post-Minkowskian approach to gravitational-wave physics and explore the feasibility
of analogous calculations in GR, we point out the possible phenomenological 
relevance of our 5PL QED amplitude as well as of amplitudes at lower PL orders~\cite{Bern:2021xze} 
to ultra-peripheral scattering as probed e.g.~at the TOTEM experiment. These aspects deserve further study.

Amplitude-based approaches efficiently solve the problem of constructing integrands 
for the foreseeable future, even in GR. The critical issue addressed here is whether the overwhelming 
increase in complexity of integrals encountered at 5PM and beyond impedes
further progress.
Electrodynamics is an especially useful test case because its diagram topologies are a subset of those appearing in GR.   
In the overlap, the integrals have identical IBP complexity in the classical limit and share the same set of master integrals. 
Moreover, most of the master integrals are also shared by more involved diagram 
topologies that appear in the gravitational calculation.
Our results indicate that the powerful field-theory integration methods are sufficient to meet the 5PM 
challenge. We anticipate much more progress to follow in the near future.

%======================================================
%
\medskip
\Section{Acknowledgements. }
%
%======================================================
%
We thank  Mikhail Solon, Chia-Hsien Shen, Anna Stasto, and Johann Usovitsch for valuable discussions and Johannes Bl\"umlein, 
Thibault Damour, Riccardo Gonzo, Anton Ilderton, and Donal O'Connell for comments on the manuscript.
We especially thank Justin Vines for discussions and making his unpublished post-Coulombian results available to us.
We are also grateful to Harald Ita for valuable discussions and for hosting the mini-workshop ``Feynman Integrals'' at ITS Zurich.
Z.B., E.H., and M.R.~are supported by the U.S. Department of Energy (DOE) under award number DE-SC0009937. 
R.R.~is supported by the U.S.  Department of Energy (DOE) under award number~DE-SC00019066.
The work of A.S. and V.S. was supported by the Russian Science
Foundation under the agreement No. 21-71-30003 (development of new
features of the \texttt{FIRE} program) and by the Ministry
of Education and Science of the Russian Federation as part of the
program of the Moscow Center for Fundamental and Applied Mathematics
under Agreement No. 075-15-2019-1621 (constructing improved
bases of master integrals).
M.Z.'s work is supported in part by the U.K.\ Royal Society through Grant 
URF\textbackslash R1\textbackslash 20109. For the purpose of open access, 
the author has applied a Creative Commons Attribution (CC BY) license to any 
Author Accepted Manuscript version arising from this submission.
We are also grateful to the Mani L. Bhaumik Institute for Theoretical Physics for support.

%%%%%%%%%%%%%%%%%%%%%%%%%%%%%%%%%%%%%%%%%%%%%%%%%%%%%%%%%%%%%%%

\end{document}

%% file: tikz_figs.tex
\tikzset{
	graviton/.style={decorate,line width=0.1mm, decoration={snake,amplitude=.5mm, segment length=2mm}},
	photon/.style={decorate, decoration={snake,amplitude=.5mm, segment length=2mm}, draw=red},
	photonBG/.style={decorate, decoration={snake,amplitude=.5mm, segment length=2mm}, draw=white,line width= 1mm},
	scalar/.style={postaction={decorate},
	},
	massive/.style={postaction={decorate},
		line width=0.75mm,
	},
	massless/.style={postaction={decorate},
	},
	masslessOverlap/.style={
		decoration={show path construction, lineto code={
				\draw[white,line width=1.5mm,line cap=round] ($(\tikzinputsegmentfirst)!0.2!(\tikzinputsegmentlast)$) to ($(\tikzinputsegmentfirst)!0.8!(\tikzinputsegmentlast)$);
				\draw (\tikzinputsegmentfirst) to (\tikzinputsegmentlast);
		}},
		decorate
	},
	masslessWithDot/.style={postaction={decorate},
		decoration={
			markings,
			mark=at position 0.5 with {\fill circle (2pt);}}
	},
	massiveWithDot/.style={postaction={decorate},
		line width=0.5mm,
		decoration={
			markings,
			mark=at position 0.5 with {\fill circle (2pt);}}
	},
	massiveLinWithDot/.style={postaction={decorate},
		double,
		thick,
		fill=white,
		decoration={
			markings,
			mark=at position 0.5 with {\fill[black] circle (2pt);}}
	},	
	massiveWithArrow/.style={postaction={decorate},
		line width=0.75mm,
		decoration={
			markings,
			mark=at position 0.5 with {\arrow{latex}}}
	},
	massiveWithArrow2/.style={postaction={decorate},
		line width=0.75mm,
		decoration={
			markings,
			mark=at position 0.9 with {\arrow{latex}}}
	},
	massiveLin/.style={postaction={decorate},
		double,
		thick,
		fill=white
	},
	massivePhi/.style={postaction={decorate},
		line width=0.75mm,
		dashed
	},
	masslessPhi/.style={postaction={decorate},
		dashed
	},
	unitaryCut/.style={postaction={draw,densely dashed,blue,thin},
		line width = 0.2cm,white
	},
	gluon/.style={decorate, draw=magenta,
		decoration={coil,amplitude=4pt, segment length=5pt}},
	partial ellipse/.style args={#1:#2:#3}{
		insert path={+ (#1:#3) arc (#1:#2:#3)}
	},
	cross/.style={cross out, draw=black, minimum size=2*(#1-\pgflinewidth), inner sep=0pt, outer sep=0pt},
	branchCut/.style={postaction={decorate},
		snake=zigzag,
		decoration = {snake=zigzag,segment length = 2mm, amplitude = 2mm}	
	},
	cross/.default={1pt},
	massiveWithResidue/.style={postaction={decorate},
		decoration={
			markings,
			mark=at position 0.5 with {\draw node[cross,red] {};}}
	},
	massiveLinWithResidue/.style={postaction={decorate},
		double,
		thick,
		fill=white,
		decoration={
			markings,
			mark=at position 0.5 with {\draw node[cross=4pt,red] {};}}	
	}
}

\newcommand{\treetPhoton}{
\begin{tikzpicture}[scale=.8]
                \coordinate (e1) at (-0.5,0.);
                \coordinate (e2) at (.5,0.);
                \coordinate (e3) at (.5,-1.);
                \coordinate (e4) at (-0.5,-1.);

                \coordinate (v1) at (0,0);
                \coordinate (v2) at (0,-1.);

                        \draw[massive] (e1) -- (v1) -- (e2) ;
                \draw[massive] (e3) -- (v2) -- (e4) ;

                \draw[photon] (v1) -- (v2) ;

                \node[below left]  at (e4) {$p_1$};
                \node[above left]  at (e1) {$p_2$};
                \node[above right] at (e2) {$p_3$};
                \node[below right] at (e3) {$p_4$};
\end{tikzpicture}}

\newcommand{\PlanarTripleLadder}{\begin{tikzpicture}[scale=0.4]
		\coordinate (e1) at (-1, -1);
		\coordinate (e2) at (-1, 1);
		\coordinate (e3) at (3.,1);
		\coordinate (e4) at (3.,-1);

		\coordinate (b1) at (-0.5, -1);
		\coordinate (b2) at (0.5, -1);
		\coordinate (b3) at (1.5, -1);
		\coordinate (b4) at (2.5, -1);
		\coordinate (t1) at (-0.5, 1);
		\coordinate (t2) at (0.5, 1);
		\coordinate (t3) at (1.5, 1);
		\coordinate (t4) at (2.5,1);
		\draw[photon] (t1) -- (b1) ;
		\draw[photon] (t2) -- (b2) ;		
		\draw[photon] (t3) -- (b3) ;
		\draw[photon] (t4) -- (b4) ;
		\draw[massive] (e1) -- (b1) -- (b2) --(b3) --(b4) --(e4) ;
		\draw[massive] (e2) -- (t1) -- (t2) --(t3) --(t4) --(e3) ;

                \node[below left]  at (e1) {$p_1$};
                \node[above left] at (e2) {$p_2$};
                \node[above right] at (e3) {$p_3$};		
                \node[below right]  at (e4) {$p_4$};
		\end{tikzpicture}
}

\newcommand{\PlanarQuadLadder}{\begin{tikzpicture}[scale=0.4]
		\coordinate (e1) at (-1, -1);
		\coordinate (e2) at (-1, 1);
		\coordinate (e3) at (4,1);
		\coordinate (e4) at (4,-1);

		\coordinate (b1) at (-0.5, -1);
		\coordinate (b2) at (0.5, -1);
		\coordinate (b3) at (1.5, -1);
		\coordinate (b4) at (2.5, -1);
		\coordinate (b5) at (3.5, -1);
		
		\coordinate (t1) at (-0.5, 1);
		\coordinate (t2) at (0.5, 1);
		\coordinate (t3) at (1.5, 1);
		\coordinate (t4) at (2.5,1);
		\coordinate (t5) at (3.5,1);

		\draw[photon] (t1) -- (b1) ;
		\draw[photon] (t2) -- (b2) ;		
		
		\draw[photon] (t3) -- (b3) ;
		\draw[photon] (t4) -- (b4) ;
		\draw[photon] (t5) -- (b5) ;

		\draw[massive] (e1) -- (b1) -- (b2) --(b3) --(b4) -- (b5) --(e4) ;
		\draw[massive] (e2) -- (t1) -- (t2) --(t3) --(t4) -- (t5) --(e3) ;
		
                \node[below left]  at (e1) {$p_1$};
                \node[above left] at (e2) {$p_2$};
                \node[above right] at (e3) {$p_3$};		
                \node[below right]  at (e4) {$p_4$};

		\end{tikzpicture}
}

%% file: 5PM_QED.bbl
\begin{thebibliography}{99}

\bibitem{LIGO}
B.~P.~Abbott {\it et al.} [LIGO Scientific and Virgo Collaborations],
%``Observation of gravitational waves from a binary black hole merger,''
Phys.\ Rev.\ Lett.\  {\bf 116}, no. 6, 061102 (2016)
%doi:10.1103/PhysRevLett.116.061102
[arXiv:1602.03837 [gr-qc]];
%%CITATION = doi:10.1103/PhysRevLett.116.061102;%% 
% 
B.~P.~Abbott {\it et al.} [LIGO Scientific and Virgo Collaborations],
%``GW170817: Observation of gravitational waves from a binary neutron star inspiral,''
Phys.\ Rev.\ Lett.\  {\bf 119}, no. 16, 161101 (2017)
%doi:10.1103/PhysRevLett.119.161101
[arXiv:1710.05832 [gr-qc]].
%%CITATION = doi:10.1103/PhysRevLett.119.161101;%%

\bibitem{NewDetectors}
M.~Punturo {\it et al.,} ``The Einstein Telescope: A third-generation gravitational wave observatory,''
 Class. Quant. Grav. 27 (2010) 194002;
%
P.~Amaro-Seoane \textit{et al.} [LISA],
%``Laser Interferometer Space Antenna,''
[arXiv:1702.00786 [astro-ph.IM]];
%844 citations counted in INSPIRE as of 05 Oct 2020
%
D.~Reitze \textit{et al.}
%``Cosmic Explorer: The U.S. Contribution to Gravitational-Wave Astronomy beyond LIGO,''
Bull. Am. Astron. Soc. \textbf{51}, 035
[arXiv:1907.04833 [astro-ph.IM]].
%120 citations counted in INSPIRE as of 30 Dec 2020
%
V.~Kalogera, B.~S.~Sathyaprakash, M.~Bailes, M.~A.~Bizouard, A.~Buonanno, A.~Burrows, M.~Colpi, M.~Evans, S.~Fairhurst and S.~Hild, \textit{et al.}
%``The Next Generation Global Gravitational Wave Observatory: The Science Book,''
[arXiv:2111.06990 [gr-qc]].
%66 citations counted in INSPIRE as of 06 May 2023

\bibitem{PN}
J.~Droste,
%``The field of $n$ moving centres in Einstein's theory of gravitation,''
Proc.\ Acad.\ Sci.\ Amst.\ 19:447–455 (1916);
%                                  
H. A. Lorentz and J. Droste, 
%``De beweging van een stelsel lichamen onder de theorie van Einstein I,II''
Koninklijke Akademie Van Wetenschappen te Amsterdam 26 392, 649 (1917).
English translation in ``Lorentz Collected papers,'' P. Zeeman and A. D. Fokker editors, Vol 5, 330 (1934-1939),
The Hague: Nijhof;
%
A.~Einstein, L.~Infeld and B.~Hoffmann,
%``The Gravitational equations and the problem of motion,''
Annals Math.\  {\bf 39}, 65 (1938).
%doi:10.2307/1968714.
%%CITATION = doi:10.2307/1968714;%%

\bibitem{PM}
B. Bertotti,
%``On  gravitational  motion'',
Nuovo  Cimento {\bf 4}, 898  (1956);
%doi:10.1007/BF02746175;
%
R. P. Kerr,
%``The Lorentz-covariant approximation method in general relativity'',
I. Nuovo Cimento {\bf 13}, 469 (1959);
%doi:10.1007/BF02732767;
%
B.~Bertotti, J.~F.~Pleba\'nski,
%``Theory of gravitational perturbations in the fast motion approximation'',
Ann, Phys, {\bf 11}, 169 (1960);
%doi:10.1016/0003-4916(60)90132-9;
%
M.~Portilla,
% ``Momentum and angular momentum of two gravitating particles,''
J.\ Phys.\ A {\bf 12}, 1075 (1979);
%doi:10.1088/0305-4470/12/7/025;
%%CITATION = doi:10.1088/0305-4470/12/7/025;%%
%
K.~Westpfahl and M.~Goller,
%``Gravitational scattering of two relativistic particles in postlinear approximation,''
Lett.\ Nuovo Cim.\  {\bf 26}, 573 (1979);
%doi:10.1007/BF02817047;
%%CITATION = doi:10.1007/BF02817047;%%
%  
M.~Portilla,
%``Scattering of two gravitating particles: classical approach,''
J.\ Phys.\ A {\bf 13}, 3677 (1980);
%doi:10.1088/0305-4470/13/12/017;
%%CITATION = doi:10.1088/0305-4470/13/12/017;%%
%
L.~Bel, T.~Damour, N.~Deruelle, J.~Ibanez and J.~Martin,
%``Poincar\'e-invariant gravitational field and equations of motion of two pointlike objects: The postlinear approximation of general relativity,''
Gen.\ Rel.\ Grav.\  {\bf 13}, 963 (1981).
% doi:10.1007/BF00756073;
%%CITATION = doi:10.1007/BF00756073;%%

\bibitem{PNPMReview}
A.~Buonanno, M.~Khalil, D.~O'Connell, R.~Roiban, M.~P.~Solon and M.~Zeng,
%``Snowmass White Paper: Gravitational Waves and Scattering Amplitudes,''
[arXiv:2204.05194 [hep-th]];
%40 citations counted in INSPIRE as of 06 May 2023}
%
G.~Sch\"afer and P.~Jaranowski,
%``Hamiltonian formulation of general relativity and post-Newtonian dynamics of compact binaries,''
Living Rev. Rel. \textbf{21}, no.1, 7 (2018)
%doi:10.1007/s41114-018-0016-5
[arXiv:1805.07240 [gr-qc]];
%96 citations counted in INSPIRE as of 10 May 2023
%
L.~Blanchet,
%``Gravitational radiation from post-Newtonian sources and inspiralling compact binaries,''
Living Rev. Rel. \textbf{9}, 4 (2006).
%427 citations counted in INSPIRE as of 07 May 2023


\bibitem{ScatteringToGrav}
Y.~Iwasaki,
%``Quantum theory of gravitation vs. classical theory---fourth-order potential,''
Prog.\ Theor.\ Phys.\  {\bf 46}, 1587 (1971);
%doi:10.1143/PTP.46.1587;\\
%%CITATION = doi:10.1143/PTP.46.1587;%%
%                                                                                                                                 
Y.~Iwasaki,
%``Fourth-order gravitational potential based on quantum field theory,''
Lett.\ Novo Cim.\  {\bf 1S2}, 783 (1971)
[Lett.\ Nuovo Cim.\  {\bf 1}, 783 (1971)];
%doi:10.1007/BF02770190;\\
%%CITATION = doi:10.1007/BF02770190;%% 
%
D.~Neill and I.~Z.~Rothstein,
%``Classical Space-Times from the S Matrix,''
Nucl. Phys. B \textbf{877}, 177-189 (2013)
%doi:10.1016/j.nuclphysb.2013.09.007
[arXiv:1304.7263 [hep-th]].
%98 citations counted in INSPIRE as of 30 Dec 2020

\bibitem{DamourPM}
T.~Damour,
%``Gravitational scattering, post-Minkowskian approximation and effective one-body theory,''
Phys.\ Rev.\ D {\bf 94}, no. 10, 104015 (2016)
%doi:10.1103/PhysRevD.94.104015
[arXiv:1609.00354 [gr-qc]];
%%CITATION = doi:10.1103/PhysRevD.94.104015;%%                                                                                    
%
T.~Damour,
%``High-energy gravitational scattering and the general relativistic two-body problem,''
Phys. Rev. D \textbf{97}, no.4, 044038 (2018)
%doi:10.1103/PhysRevD.97.044038
[arXiv:1710.10599 [gr-qc]].
%210 citations counted in INSPIRE as of 15 Apr 2023

\bibitem{BCJ}
H.~Kawai, D.~C.~Lewellen and S.~H.~H.~Tye,
%``A relation between tree amplitudes of closed and open strings,''
Nucl.\ Phys.\ B {\bf 269}, 1 (1986);
%doi:10.1016/0550-3213(86)90362-7;
%%CITATION = doi:10.1016/0550-3213(86)90362-7;%%
%
Z.~Bern, L.~J.~Dixon, M.~Perelstein and J.~S.~Rozowsky,
%``Multileg one loop gravity amplitudes from gauge theory,''
Nucl.\ Phys.\ B {\bf 546}, 423 (1999)
%doi:10.1016/S0550-3213(99)00029-2
[hep-th/9811140];
%%CITATION = doi:10.1016/S0550-3213(99)00029-2;%%
%
Z.~Bern, J.~J.~M.~Carrasco and H.~Johansson,
%``New relations for gauge-theory amplitudes,''
Phys.\ Rev.\ D {\bf 78}, 085011 (2008)
%doi:10.1103/PhysRevD.78.085011
[arXiv:0805.3993 [hep-ph]];
%%CITATION = doi:10.1103/PhysRevD.78.085011;%%                                                             
%                                                                                                          
Z.~Bern, J.~J.~M.~Carrasco and H.~Johansson,
%``Perturbative quantum gravity as a double copy of gauge theory,''
Phys.\ Rev.\ Lett.\  {\bf 105}, 061602 (2010)
%doi:10.1103/PhysRevLett.105.061602                                                                        
[arXiv:1004.0476 [hep-th]];
%%CITATION = doi:10.1103/PhysRevLett.105.061602;%%
%                                                                                                          
%
Z.~Bern, J.~J.~Carrasco, M.~Chiodaroli, H.~Johansson and R.~Roiban,
%``The duality between color and kinematics and its applications,''
arXiv:1909.01358 [hep-th].
%%CITATION = ARXIV:1909.01358;%%


\bibitem{GeneralizedUnitarity}
Z.~Bern, L.~J.~Dixon, D.~C.~Dunbar and D.~A.~Kosower,
%``One loop $n$-point gauge-theory amplitudes, unitarity and collinear limits,''
Nucl.\ Phys.\ B {\bf 425}, 217 (1994)
%doi:10.1016/0550-3213(94)90179-1
[hep-ph/9403226];
%%CITATION = doi:10.1016/0550-3213(94)90179-1;%%
%
Z.~Bern, L.~J.~Dixon, D.~C.~Dunbar and D.~A.~Kosower,
%``Fusing gauge-theory tree amplitudes into loop amplitudes,''
Nucl.\ Phys.\ B {\bf 435}, 59 (1995)
%doi:10.1016/0550-3213(94)00488-Z
[hep-ph/9409265];
%%CITATION = doi:10.1016/0550-3213(94)00488-Z;%%
%
Z.~Bern and A.~G.~Morgan,
%``Massive loop amplitudes from unitarity,''
Nucl.\ Phys.\ B {\bf 467}, 479 (1996)
%doi:10.1016/0550-3213(96)00078-8
[hep-ph/9511336];
%%CITATION = doi:10.1016/0550-3213(96)00078-8;%%
%
Z.~Bern, L.~J.~Dixon and D.~A.~Kosower,
%``One loop amplitudes for $e^+ e^-$ to four partons,''
Nucl.\ Phys.\ B {\bf 513}, 3 (1998)
%doi:10.1016/S0550-3213(97)00703-7
[hep-ph/9708239];
%%CITATION = doi:10.1016/S0550-3213(97)00703-7;%%
%
R.~Britto, F.~Cachazo and B.~Feng,
%``Generalized unitarity and one-loop amplitudes in $N=4$ super-Yang-Mills,''
Nucl.\ Phys.\ B {\bf 725}, 275 (2005)
%doi:10.1016/j.nuclphysb.2005.07.014
[hep-th/0412103];
%%CITATION = doi:10.1016/j.nuclphysb.2005.07.014;%%
%
Z.~Bern, J.~J.~M.~Carrasco, H.~Johansson and D.~A.~Kosower,
%``Maximally supersymmetric planar Yang-Mills amplitudes at five loops,''
Phys.\ Rev.\ D {\bf 76}, 125020 (2007)
%doi:10.1103/PhysRevD.76.125020
[arXiv:0705.1864 [hep-th]].
%%CITATION = doi:10.1103/PhysRevD.76.125020;%%

\bibitem{IBP}
K.G.~Chetyrkin and F.V.~Tkachov,
%``Integration by parts: the algorithm to calculate beta functions in 4 loops,''
Nucl.\ Phys.\ B {\bf 192}, 159 (1981);
%%CITATION = NUPHA,B192,159;%% 
%
S.~Laporta,
%``High precision calculation of multiloop Feynman integrals by difference equations,''
Int. J. Mod. Phys. A \textbf{15}, 5087-5159 (2000)
%doi:10.1016/S0217-751X(00)00215-7
[arXiv:hep-ph/0102033 [hep-ph]];
%
A.~V.~Smirnov and V.~A.~Smirnov,
%``How to choose master integrals,''
Nucl. Phys. B \textbf{960}, 115213 (2020)
%doi:10.1016/j.nuclphysb.2020.115213
[arXiv:2002.08042 [hep-ph]];
%42 citations counted in INSPIRE as of 14 May 2023
%
J.~Usovitsch,
%``Factorization of denominators in integration-by-parts reductions,''
[arXiv:2002.08173 [hep-ph]].
%33 citations counted in INSPIRE as of 14 May 2023

\bibitem{DEs}
A.~V.~Kotikov,
%``Differential equations method: New technique for massive Feynman diagrams calculation,''
Phys. Lett. B \textbf{254}, 158-164 (1991);
%doi:10.1016/0370-2693(91)90413-K
%590 citations counted in INSPIRE as of 15 Oct 2020
%
Z.~Bern, L.~J.~Dixon and D.~A.~Kosower,
%``Dimensionally regulated pentagon integrals,''
Nucl. Phys. B \textbf{412}, 751-816 (1994)
%doi:10.1016/0550-3213(94)90398-0
[arXiv:hep-ph/9306240 [hep-ph]];
%487 citations counted in INSPIRE as of 05 Oct 2020
%
E.~Remiddi,
%``Differential equations for Feynman graph amplitudes,''
Nuovo Cim. A \textbf{110}, 1435-1452 (1997)
[arXiv:hep-th/9711188 [hep-th]];
%474 citations counted in INSPIRE as of 15 Oct 2020
%
T.~Gehrmann and E.~Remiddi,
%``Differential equations for two loop four point functions,''
Nucl. Phys. B \textbf{580}, 485-518 (2000)
%doi:10.1016/S0550-3213(00)00223-6
[arXiv:hep-ph/9912329 [hep-ph]].
%642 citations counted in INSPIRE as of 05 Oct 2020

\bibitem{Henn} 
J.~M.~Henn,
%``Multiloop integrals in dimensional regularization made simple,''
Phys. Rev. Lett. \textbf{110}, 251601 (2013)
%doi:10.1103/PhysRevLett.110.251601
[arXiv:1304.1806 [hep-th]];
%485 citations counted in INSPIRE as of 10 Jan 2021
%
J.~M.~Henn, A.~V.~Smirnov and V.~A.~Smirnov,
%``Evaluating single-scale and/or non-planar diagrams by differential equations,''
JHEP \textbf{03}, 088 (2014)
%doi:10.1007/JHEP03(2014)088
[arXiv:1312.2588 [hep-th]].
%93 citations counted in INSPIRE as of 10 Jan 2021

\bibitem{NRGR}
W.~D.~Goldberger and I.~Z.~Rothstein,
%``An effective field theory of gravity for extended objects,''
Phys.\ Rev.\ D {\bf 73}, 104029 (2006)
%doi:10.1103/PhysRevD.73.104029
[hep-th/0409156].
%%CITATION = doi:10.1103/PhysRevD.73.104029;%%

\bibitem{2PM}
C.~Cheung, I.~Z.~Rothstein and M.~P.~Solon,
%``From Scattering Amplitudes to Classical Potentials in the Post-Minkowskian Expansion,''
Phys. Rev. Lett. \textbf{121}, no.25, 251101 (2018)
%doi:10.1103/PhysRevLett.121.251101
[arXiv:1808.02489 [hep-th]].
%121 citations counted in INSPIRE as of 27 Dec 2020

\bibitem{3PMPRL}
Z.~Bern, C.~Cheung, R.~Roiban, C.~H.~Shen, M.~P.~Solon and M.~Zeng,
%``Scattering Amplitudes and the Conservative Hamiltonian for Binary Systems at Third Post-Minkowskian Order,''
Phys. Rev. Lett. \textbf{122}, no.20, 201603 (2019)
%doi:10.1103/PhysRevLett.122.201603
[arXiv:1901.04424 [hep-th]].
%139 citations counted in INSPIRE as of 27 Dec 2020

\bibitem{3PMLong}
Z.~Bern, C.~Cheung, R.~Roiban, C.~H.~Shen, M.~P.~Solon and M.~Zeng,
%``Black Hole Binary Dynamics from the Double Copy and Effective Theory,''
JHEP \textbf{10}, 206 (2019)
%doi:10.1007/JHEP10(2019)206
[arXiv:1908.01493 [hep-th]].
%112 citations counted in INSPIRE as of 27 Dec 2020

\bibitem{4PMPotential}
Z.~Bern, J.~Parra-Martinez, R.~Roiban, M.~S.~Ruf, C.~H.~Shen, M.~P.~Solon and M.~Zeng,
%``Scattering Amplitudes and Conservative Binary Dynamics at ${\cal O}(G^4)$,''
Phys. Rev. Lett. \textbf{126}, no.17, 171601 (2021)
%doi:10.1103/PhysRevLett.126.171601
[arXiv:2101.07254 [hep-th]].
%68 citations counted in INSPIRE as of 02 Dec 2021

\bibitem{4PMTail}
Z.~Bern, J.~Parra-Martinez, R.~Roiban, M.~S.~Ruf, C.~H.~Shen, M.~P.~Solon and M.~Zeng,
%``Scattering Amplitudes, the Tail Effect, and Conservative Binary Dynamics at O(G4),''
Phys. Rev. Lett. \textbf{128}, no.16, 161103 (2022)
%doi:10.1103/PhysRevLett.128.161103
[arXiv:2112.10750 [hep-th]].
%68 citations counted in INSPIRE as of 13 Apr 2023


\bibitem{Dlapa:2021npj}
C.~Dlapa, G.~K\"alin, Z.~Liu and R.~A.~Porto,
%``Dynamics of Binary Systems to Fourth Post-Minkowskian Order from the Effective Field Theory Approach,''
Phys. Lett. B \textbf{831}, 137203 (2022)
%doi:10.1016/j.physletb.2022.137203
[arXiv:2106.08276 [hep-th]].
%92 citations counted in INSPIRE as of 15 Apr 2023

\bibitem{Dlapa:2021vgp}
C.~Dlapa, G.~K\"alin, Z.~Liu and R.~A.~Porto,
%``Conservative Dynamics of Binary Systems at Fourth Post-Minkowskian Order in the Large-Eccentricity Expansion,''
Phys. Rev. Lett. \textbf{128}, no.16, 161104 (2022)
%doi:10.1103/PhysRevLett.128.161104
[arXiv:2112.11296 [hep-th]].
%52 citations counted in INSPIRE as of 15 Apr 2023

\bibitem{Manohar:2022dea}
A.~V.~Manohar, A.~K.~Ridgway and C.~H.~Shen,
%``Radiated Angular Momentum and Dissipative Effects in Classical Scattering,''
Phys. Rev. Lett. \textbf{129}, no.12, 121601 (2022)
%doi:10.1103/PhysRevLett.129.121601
[arXiv:2203.04283 [hep-th]].
%37 citations counted in INSPIRE as of 15 Apr 2023

\bibitem{Dlapa:2022lmu}
C.~Dlapa, G.~K\"alin, Z.~Liu, J.~Neef and R.~A.~Porto,
%``Radiation Reaction and Gravitational Waves at Fourth Post-Minkowskian Order,''
Phys. Rev. Lett. \textbf{130}, no.10, 101401 (2023)
%doi:10.1103/PhysRevLett.130.101401
[arXiv:2210.05541 [hep-th]].
%22 citations counted in INSPIRE as of 15 Apr 2023

\bibitem{NR}
F.~Pretorius,
%``Evolution of binary black hole spacetimes,''
Phys.\ Rev.\ Lett.\  {\bf 95}, 121101 (2005)
%doi:10.1103/PhysRevLett.95.121101
[gr-qc/0507014];
%%CITATION = doi:10.1103/PhysRevLett.95.121101;%%
%
M.~Campanelli, C.~O.~Lousto, P.~Marronetti and Y.~Zlochower,
%``Accurate evolutions of orbiting black-hole binaries without excision,''
Phys.\ Rev.\ Lett.\  {\bf 96}, 111101 (2006)
%doi:10.1103/PhysRevLett.96.111101
[gr-qc/0511048];
%%CITATION = doi:10.1103/PhysRevLett.96.111101;%%
%
J.~G.~Baker, J.~Centrella, D.~I.~Choi, M.~Koppitz and J.~van Meter,
%``Gravitational wave extraction from an inspiraling configuration of merging black holes,''
Phys.\ Rev.\ Lett.\  {\bf 96}, 111102 (2006)
%doi:10.1103/PhysRevLett.96.111102
[gr-qc/0511103].
%%CITATION = doi:10.1103/PhysRevLett.96.111102;%%

\bibitem{self_force} 
Y.~Mino, M.~Sasaki and T.~Tanaka,
%``Gravitational radiation reaction to a particle motion,''
Phys.\ Rev.\ D {\bf 55}, 3457 (1997)
%doi:10.1103/PhysRevD.55.3457
[gr-qc/9606018];
%%CITATION = doi:10.1103/PhysRevD.55.3457;%%
%
T.~C.~Quinn and R.~M.~Wald,
%``An Axiomatic approach to electromagnetic and gravitational radiation reaction of particles in curved space-time,''
Phys.\ Rev.\ D {\bf 56}, 3381 (1997)
%doi:10.1103/PhysRevD.56.3381
[gr-qc/9610053];
%%CITATION = doi:10.1103/PhysRevD.56.3381;%%
%  one review
L.~Barack and A.~Pound,
%``Self-force and radiation reaction in general relativity,''
Rept. Prog. Phys. \textbf{82}, no.1, 016904 (2019)
%doi:10.1088/1361-6633/aae552
[arXiv:1805.10385 [gr-qc]].


\bibitem{EOB}
A.~Buonanno and T.~Damour,
%``Effective one-body approach to general relativistic two-body dynamics,''
Phys.\ Rev.\ D {\bf 59}, 084006 (1999)
%doi:10.1103/PhysRevD.59.084006
[gr-qc/9811091];
%%CITATION = doi:10.1103/PhysRevD.59.084006;%% 
%
A.~Buonanno and T.~Damour,
%``Transition from inspiral to plunge in binary black hole coalescences,''
Phys.\ Rev.\ D {\bf 62}, 064015 (2000)
%doi:10.1103/PhysRevD.62.064015
[gr-qc/0001013].
%%CITATION = doi:10.1103/PhysRevD.62.064015;%%

\bibitem{Khalil:2022ylj}
M.~Khalil, A.~Buonanno, J.~Steinhoff and J.~Vines,
%``Energetics and scattering of gravitational two-body systems at fourth post-Minkowskian order,''
Phys. Rev. D \textbf{106}, no.2, 024042 (2022)
%doi:10.1103/PhysRevD.106.024042
[arXiv:2204.05047 [gr-qc]];
%
%\bibitem{Damour:2022ybd}
T.~Damour and P.~Rettegno,
%``Strong-field scattering of two black holes: Numerical relativity meets post-Minkowskian gravity,''
Phys. Rev. D \textbf{107}, no.6, 064051 (2023)
%doi:10.1103/PhysRevD.107.064051
[arXiv:2211.01399 [gr-qc]].
%8 citations counted in INSPIRE as of 12 Apr 2023

\bibitem{FiveLoops}
Z.~Bern, J.~J.~M.~Carrasco, W.~M.~Chen, H.~Johansson, R.~Roiban and M.~Zeng,
%``Five-loop four-point integrand of $N=8$ supergravity as a generalized double copy,''
Phys. Rev. D \textbf{96}, no.12, 126012 (2017)
%doi:10.1103/PhysRevD.96.126012
[arXiv:1708.06807 [hep-th]].

\bibitem{Westpfhal}
K. Westpfahl,
%``High-speed scattering of charged and uncharged particles in general relativity'',
Fortschr. Phys., {\bf 33}, 417 (1985).
 
\bibitem{Damour:1990jh}
T.~Damour and G.~Schaefer,
%``Redefinition of position variables and the reduction of higher order Lagrangians,''
J. Math. Phys. \textbf{32}, 127-134 (1991).
%doi:10.1063/1.529135
%87 citations counted in INSPIRE as of 10 May 2023


\bibitem{Buonanno:2000qq}
A.~Buonanno,
%``Reduction of the two-body dynamics to a one-body description in classical electrodynamics,''
Phys. Rev. D \textbf{62}, 104022 (2000)
%doi:10.1103/PhysRevD.62.104022
[arXiv:hep-th/0004042 [hep-th]].
%11 citations counted in INSPIRE as of 24 Apr 2023


\bibitem{KMOC}
D.~A.~Kosower, B.~Maybee and D.~O'Connell,
%``Amplitudes, Observables, and Classical Scattering,''
JHEP \textbf{02}, 137 (2019)
%doi:10.1007/JHEP02(2019)137
[arXiv:1811.10950 [hep-th]].

\bibitem{Saketh:2021sri}
M.~V.~S.~Saketh, J.~Vines, J.~Steinhoff and A.~Buonanno,
%``Conservative and radiative dynamics in classical relativistic scattering and bound systems,''
Phys. Rev. Res. \textbf{4}, no.1, 013127 (2022)
%doi:10.1103/PhysRevResearch.4.013127
[arXiv:2109.05994 [gr-qc]].
%36 citations counted in INSPIRE as of 12 Apr 2023

\bibitem{Bern:2021xze}
Z.~Bern, J.~P.~Gatica, E.~Herrmann, A.~Luna and M.~Zeng,
%``Scalar QED as a toy model for higher-order effects in classical gravitational scattering,''
JHEP \textbf{08}, 131 (2022)
%doi:10.1007/JHEP08(2022)131
[arXiv:2112.12243 [hep-th]].
%15 citations counted in INSPIRE as of 12 Apr 2023


\bibitem{FIRE}
A.~V.~Smirnov,
%``Algorithm FIRE -- Feynman Integral REduction,''
JHEP \textbf{10}, 107 (2008)
%doi:10.1088/1126-6708/2008/10/107
[arXiv:0807.3243 [hep-ph]];
%427 citations counted in INSPIRE as of 30 Sep 2020
%
A.~V.~Smirnov,
%``FIRE5: a C++ implementation of Feynman Integral REduction,''
Comput. Phys. Commun. \textbf{189}, 182-191 (2015)
%doi:10.1016/j.cpc.2014.11.024
[arXiv:1408.2372 [hep-ph]];
%397 citations counted in INSPIRE as of 07 May 2023
%
A.~V.~Smirnov and F.~S.~Chuharev,
%``FIRE6: Feynman Integral REduction with Modular Arithmetic,''
%doi:10.1016/j.cpc.2019.106877
[arXiv:1901.07808 [hep-ph]].
%64 citations counted in INSPIRE as of 30 Sep 2020 

\bibitem{Lee:2013mka}
R.~N.~Lee,
%``LiteRed 1.4: a powerful tool for reduction of multiloop integrals,''
J. Phys. Conf. Ser. \textbf{523}, 012059 (2014)
%doi:10.1088/1742-6596/523/1/012059
[arXiv:1310.1145 [hep-ph]].
%444 citations counted in INSPIRE as of 29 Apr 2023

\bibitem{TailEffect}
W.~Bonnor, Philos. Trans. R. Soc. London, Ser. A 251, 233 (1959);
%                                                                                                                    
W.~Bonnor and M.~Rotenberg, Proc. R. Soc. London, Ser. A 289, 247 (1966);
%
K.~S.~Thorne,
%``Multipole expansions of gravitational radiation,''
Rev.\ Mod.\ Phys.\  {\bf 52}, 299 (1980);
%doi:10.1103/RevModPhys.52.299;
%%CITATION = doi:10.1103/RevModPhys.52.299;%%
%
L.~Blanchet and T.~Damour,
%``Tail transported temporal correlations in the dynamics of a gravitating system,''
Phys.\ Rev.\ D {\bf 37}, 1410 (1988);
%doi:10.1103/PhysRevD.37.1410;
%%CITATION = doi:10.1103/PhysRevD.37.1410;%%
%
L.~Blanchet and T.~Damour,
%``Hereditary effects in gravitational radiation,''
Phys.\ Rev.\ D {\bf 46}, 4304 (1992);
%doi:10.1103/PhysRevD.46.4304;
%%CITATION = doi:10.1103/PhysRevD.46.4304;%%
%
L.~Blanchet and G.~Schaefer,
%``Gravitational wave tails and binary star systems,''
Class. Quant. Grav. \textbf{10}, 2699 (1993).
%doi:10.1088/0264-9381/10/12/026
%137 citations counted in INSPIRE as of 15 Jan 2021

\bibitem{TOTEM:2017sdy}
G.~Antchev \textit{et al.} [TOTEM],
%``First determination of the ${\rho }$ parameter at ${\sqrt{s} = 13}$ TeV: probing the existence of a colourless C-odd three-gluon compound state,''
Eur. Phys. J. C \textbf{79}, no.9, 785 (2019)
%doi:10.1140/epjc/s10052-019-7223-4
[arXiv:1812.04732 [hep-ex]].
%172 citations counted in INSPIRE as of 26 Apr 2023

\bibitem{Bethe:1958zz}
H.~A.~Bethe,
%``Scattering and polarization of protons by nuclei,''
Annals Phys. \textbf{3}, 190 (1958)
%doi:10.1016/0003-4916(58)90017-4
%221 citations counted in INSPIRE as of 26 Apr 2023

\bibitem{Islam:1967zz}
M.~M.~Islam,
%``Bethe's Formula for Coulomb-Nuclear Interference,''
Phys. Rev. \textbf{162}, 1426 (1967);
%doi:10.1103/PhysRev.162.1426;
%29 citations counted in INSPIRE as of 26 Apr 2023
%
%\bibitem{West:1968du}
G.~B.~West and D.~R.~Yennie,
%``Coulomb interference in high-energy scattering,''
Phys. Rev. \textbf{172}, 1413 (1968);
%doi:10.1103/PhysRev.172.1413
%246 citations counted in INSPIRE as of 26 Apr 2023
%
%\bibitem{Cahn:1982nr}
R.~Cahn,
%``Coulombic - Hadronic Interference in an Eikonal Model,''
Z. Phys. C \textbf{15}, 253 (1982);
%doi:10.1007/BF01475009
%127 citations counted in INSPIRE as of 26 Apr 2023
%
%\bibitem{Buttimore:1978ry}
N.~H.~Buttimore, E.~Gotsman and E.~Leader,
%``SPIN DEPENDENT PHENOMENA INDUCED BY ELECTROMAGNETIC HADRONIC INTERFERENCE AT HIGH-ENERGIES,''
Phys. Rev. D \textbf{18}, 694-716 (1978)
[erratum: Phys. Rev. D \textbf{35}, no.1, 407 (1987)].
%doi:10.1103/PhysRevD.18.694
%130 citations counted in INSPIRE as of 26 Apr 2023

\bibitem{Kaspar:2011eva}
J.~Kaspar,
%``Elastic scattering at the LHC,''
CERN-THESIS-2011-214.
%13 citations counted in INSPIRE as of 26 Apr 2023

\bibitem{Khoze:2019fhx}
V.~A.~Khoze, A.~D.~Martin and M.~G.~Ryskin,
%``Bethe phase including proton excitations,''
Phys. Rev. D \textbf{101}, no.1, 016018 (2020)
%doi:10.1103/PhysRevD.101.016018
[arXiv:1910.03533 [hep-ph]].
%4 citations counted in INSPIRE as of 26 Apr 2023

\bibitem{Kaspar:2020oih}
J.~Ka\v{s}par,
%``Coulomb-nuclear Interference in Elastic Scattering: Eikonal Calculation to All Orders of $\alpha$,''
Acta Phys. Polon. B \textbf{52}, no.2, 85-97 (2021)
%doi:10.5506/APhysPolB.52.85
[arXiv:2001.10227 [hep-ph]].
%5 citations counted in INSPIRE as of 26 Apr 2023

\bibitem{Petrov:2022fsu}
V.~A.~Petrov and N.~P.~Tkachenko,
%``Coulomb-nuclear interference: Theory and practice for pp-scattering at 13~TeV,''
Phys. Rev. D \textbf{106}, no.5, 054003 (2022)
%doi:10.1103/PhysRevD.106.054003
[arXiv:2204.08815 [hep-ph]].
%2 citations counted in INSPIRE as of 26 Apr 2023

\bibitem{Petrov:2020tnr}
V.~A.~Petrov,
%``Coulomb-Nuclear Interference:the Latest Modification,''
Proc. Steklov Inst. Math. \textbf{309}, no.1, 219-224 (2020)
%doi:10.1134/S0081543820030165
[arXiv:2001.06220 [hep-ph]].
%7 citations counted in INSPIRE as of 26 Apr 2023

\bibitem{OtherApproaches}
P.~H.~Damgaard, L.~Plante and P.~Vanhove,
%``On an exponential representation of the gravitational S-matrix,''
JHEP \textbf{11}, 213 (2021)
%doi:10.1007/JHEP11(2021)213
[arXiv:2107.12891 [hep-th]];
%55 citations counted in INSPIRE as of 02 May 2023
%
A.~Brandhuber, G.~Chen, G.~Travaglini and C.~Wen,
%``Classical gravitational scattering from a gauge-invariant double copy,''
JHEP \textbf{10}, 118 (2021)
%doi:10.1007/JHEP10(2021)118
[arXiv:2108.04216 [hep-th]];
%93 citations counted in INSPIRE as of 02 May 2023
%
P.~Di Vecchia, C.~Heissenberg, R.~Russo and G.~Veneziano,
%``Classical Gravitational Observables from the Eikonal Operator,''
[arXiv:2210.12118 [hep-th]].
%12 citations counted in INSPIRE as of 02 May 2023

\bibitem{Landau:1975pou}
L.~D.~Landau and E.~M.~Lifshitz,
%``The Classical Theory of Fields,''
Pergamon Press, 1975,
ISBN 978-0-08-018176-9
%88 citations counted in INSPIRE as of 15 May 2023

\bibitem{Beneke:1997zp}
M.~Beneke and V.~A.~Smirnov,
%``Asymptotic expansion of Feynman integrals near threshold,''
Nucl. Phys. B \textbf{522}, 321-344 (1998)
%doi:10.1016/S0550-3213(98)00138-2
[arXiv:hep-ph/9711391 [hep-ph]].
%667 citations counted in INSPIRE as of 11 Jan 2021

\bibitem{Kerner}
E.~H.~Kerner, J. Math. Phys. \textbf{3}, 35 (1962).
% Hamiltonian formulation of action-at-a-distance in electrodynamics
%https://pubs.aip.org/aip/jmp/article/3/1/35/453757/Hamiltonian-Formulation-of-Action-at-a-Distance-in

\bibitem{PRZ}
J.~Parra-Martinez, M.~S.~Ruf and M.~Zeng,
%``Extremal black hole scattering at $\mathcal{O}(G^3)$: graviton dominance, eikonal exponentiation, and differential equations,''
JHEP \textbf{11}, 023 (2020)
%doi:10.1007/JHEP11(2020)023
[arXiv:2005.04236 [hep-th]].

\bibitem{Lee:2014ioa}
R.~N.~Lee,
%``Reducing differential equations for multiloop master integrals,''
JHEP \textbf{04}, 108 (2015)
%doi:10.1007/JHEP04(2015)108
[arXiv:1411.0911 [hep-ph]].
%221 citations counted in INSPIRE as of 10 Apr 2023

\bibitem{Prausa:2017ltv}
M.~Prausa,
%``epsilon: A tool to find a canonical basis of master integrals,''
Comput. Phys. Commun. \textbf{219}, 361-376 (2017)
%doi:10.1016/j.cpc.2017.05.026
[arXiv:1701.00725 [hep-ph]];
%107 citations counted in INSPIRE as of 10 Apr 2023
%\bibitem{Peraro:2019svx}
T.~Peraro,
%``FiniteFlow: multivariate functional reconstruction using finite fields and dataflow graphs,''
JHEP \textbf{07}, 031 (2019)
%doi:10.1007/JHEP07(2019)031
[arXiv:1905.08019 [hep-ph]];
%125 citations counted in INSPIRE as of 10 Apr 2023
%
%\bibitem{Dlapa:2020cwj}
C.~Dlapa, J.~Henn and K.~Yan,
%``Deriving canonical differential equations for Feynman integrals from a single uniform weight integral,''
JHEP \textbf{05}, 025 (2020)
%doi:10.1007/JHEP05(2020)025
[arXiv:2002.02340 [hep-ph]].
%52 citations counted in INSPIRE as of 10 Apr 2023

\bibitem{Goncharov:1998kja}
A.~B.~Goncharov,
%``Multiple polylogarithms, cyclotomy and modular complexes,‘’
Math. Res. Lett. \textbf{5}, 497-516 (1998)
%doi:10.4310/MRL.1998.v5.n4.a7
[arXiv:1105.2076 [math.AG]];
%485 citations counted in INSPIRE as of 05 May 2023 (edited) 
%
A.~B.~Goncharov,
%``Multiple polylogarithms and mixed Tate motives,‘’
[arXiv:math/0103059 [math.AG]].
%288 citations counted in INSPIRE as of 05 May 2023

\bibitem{AttachedFile}
See the ancillary files of this manuscript.

\bibitem{3PMEnergyLoss}
D.~Bini and T.~Damour,
%``Gravitational scattering of two black holes at the fourth post-Newtonian approximation,''
Phys. Rev. D \textbf{96}, no.6, 064021 (2017)
%doi:10.1103/PhysRevD.96.064021
[arXiv:1706.06877 [gr-qc]];
%32 citations counted in INSPIRE as of 16 Jan 2021
%
L.~Blanchet, S.~Foffa, F.~Larrouturou and R.~Sturani,
%``Logarithmic tail contributions to the energy function of circular compact binaries,''
Phys. Rev. D \textbf{101}, no.8, 084045 (2020)
%doi:10.1103/PhysRevD.101.084045
[arXiv:1912.12359 [gr-qc]];
%
D.~Bini, T.~Damour and A.~Geralico,
%``Sixth post-Newtonian nonlocal-in-time dynamics of binary systems,''
Phys. Rev. D \textbf{102}, no.8, 084047 (2020)
%doi:10.1103/PhysRevD.102.084047
[arXiv:2007.11239 [gr-qc]].
%8 citations counted in INSPIRE as of 28 Dec 2020

\bibitem{Ablinger:2011te}
J.~Ablinger, J.~Blumlein and C.~Schneider,
%``Harmonic Sums and Polylogarithms Generated by Cyclotomic Polynomials,''
J. Math. Phys. \textbf{52}, 102301 (2011)
%doi:10.1063/1.3629472
[arXiv:1105.6063 [math-ph]].
%249 citations counted in INSPIRE as of 10 Apr 2023

\bibitem{Ablinger:2018zwz}
J.~Ablinger, J.~Bl\"umlein, P.~Marquard, N.~Rana and C.~Schneider,
%``Automated Solution of First Order Factorizable Systems of Differential Equations in One Variable,''
Nucl. Phys. B \textbf{939}, 253-291 (2019)
doi:10.1016/j.nuclphysb.2018.12.010
[arXiv:1810.12261 [hep-ph]].
%64 citations counted in INSPIRE as of 15 May 2023

%\cite{Landau:1975pou}
\bibitem{VinesUnpublished}
J. Vines, unpublished.

\bibitem{QEDProbe}
C.~G.~Darwin,
%  ``‘‘On some orbits of an electron,’’ 
The London, Edinburgh, and Dublin Philosophical Magazine and Journal of Science, \textbf{25}, 201 (1913) no.146;
%doi: 10.1080/14786440208634017;
%T.~H.~Boyer,
%``Unfamiliar trajectories for a relativistic particle in a Kepler or Coulomb potential''
%American Journal of Physics, \textbf{72}, 992 (2004)
%https://doi.org/10.1119/1.1737396
%doi:10.1119/1.1737396

\bibitem{B2B}
G.~K\"alin and R.~A.~Porto,
%``From Boundary Data to Bound States,''
JHEP \textbf{01}, 072 (2020)
%doi:10.1007/JHEP01(2020)072
[arXiv:1910.03008 [hep-th]];
%54 citations counted in INSPIRE as of 08 Jan 2021
%
G.~K\"alin and R.~A.~Porto,
%``From boundary data to bound states. Part II. Scattering angle to dynamical invariants (with twist),''
JHEP \textbf{02}, 120 (2020)
%doi:10.1007/JHEP02(2020)120
[arXiv:1911.09130 [hep-th]].
%34 citations counted in INSPIRE as of 08 Jan 2021

%\cite{Ablinger:2018zwz}
\end{thebibliography}
